\documentclass[prb,amsmath,amssymb,superscriptaddress,twocolumn]{revtex4}
\usepackage{float}
\usepackage{amsmath}

\usepackage{amssymb}
\usepackage{amsfonts}
\usepackage{euscript}
\usepackage{enumerate}
\usepackage{hhline}
\usepackage{pslatex}
\usepackage{tabularx}
\usepackage[usenames,dvipsnames]{xcolor}

\usepackage{graphicx}

\usepackage{dcolumn}
\usepackage{bm}
\usepackage[sort&compress]{natbib}

\makeatletter
\renewcommand{\@biblabel}[1]{#1. }
\renewcommand{\@dotsep}{500}
\renewcommand{\@pnumwidth}{0em}
\renewcommand{\l@figure}[2]{
\@dottedtocline{1}{1.5em}{2em}{Figure #1}{}\vspace{15pt}}

\newcommand{\kst}[1]{\textcolor{black}{#1}}

\newcommand{\xl}[1]{\textcolor{black}{#1}}

\usepackage[normalem]{ulem}

\begin{document}

\title{Hybrid-mode-family Kerr optical parametric oscillation for robust coherent light generation on chip}

\author{Feng Zhou}
\affiliation{Microsystems and Nanotechnology Division, Physical Measurement Laboratory, National Institute of Standards and Technology, Gaithersburg, MD 20899, USA}
\affiliation{Theiss Research, La Jolla, California 92037, USA}
\author{Xiyuan Lu}  \email{xiyuan.lu@nist.gov}
\affiliation{Microsystems and Nanotechnology Division, Physical Measurement Laboratory, National Institute of Standards and Technology, Gaithersburg, MD 20899, USA}
\affiliation{Institute for Research in Electronics and Applied Physics and Maryland NanoCenter, University of Maryland,
College Park, MD 20742, USA}
\author{Ashutosh Rao}
\affiliation{Microsystems and Nanotechnology Division, Physical Measurement Laboratory, National Institute of Standards and Technology, Gaithersburg, MD 20899, USA}
\affiliation{Institute for Research in Electronics and Applied Physics and Maryland NanoCenter, University of Maryland,
College Park, MD 20742, USA}
\author{Jordan Stone}
\affiliation{Microsystems and Nanotechnology Division, Physical Measurement Laboratory, National Institute of Standards and Technology, Gaithersburg, MD 20899, USA}
\affiliation{Joint Quantum Institute, NIST/University of Maryland,
College Park, MD 20742, USA}
\author{Gregory Moille}
\affiliation{Microsystems and Nanotechnology Division, Physical Measurement Laboratory, National Institute of Standards and Technology, Gaithersburg, MD 20899, USA}
\affiliation{Joint Quantum Institute, NIST/University of Maryland,
College Park, MD 20742, USA}
\author{Edgar Perez}
\affiliation{Microsystems and Nanotechnology Division, Physical Measurement Laboratory, National Institute of Standards and Technology, Gaithersburg, MD 20899, USA}
\affiliation{Joint Quantum Institute, NIST/University of Maryland,
College Park, MD 20742, USA}
\author{Daron Westly}
\affiliation{Microsystems and Nanotechnology Division, Physical Measurement Laboratory, National Institute of Standards and Technology, Gaithersburg, MD 20899, USA}
\author{Kartik Srinivasan} \email{kartik.srinivasan@nist.gov}
\affiliation{Microsystems and Nanotechnology Division, Physical Measurement Laboratory, National Institute of Standards and Technology, Gaithersburg, MD 20899, USA}
\affiliation{Joint Quantum Institute, NIST/University of Maryland, College Park, MD 20742, USA}
\date{\today}

\begin{abstract}
     Optical parametric oscillation (OPO) using the third-order nonlinearity ($\chi^{(3)}$) in integrated photonics platforms is an emerging approach for coherent light generation, and has shown great promise in achieving broad spectral coverage with small device footprints and at low pump powers. However, current $\chi^{(3)}$ nanophotonic OPO devices use pump, signal, and idler modes of the same transverse spatial mode family. As a result, such single-mode-family OPO (sOPO) is inherently sensitive in dispersion and can be challenging to scalably fabricate and implement. In this work, we propose to use different families of transverse spatial modes for pump, signal, and idler, which we term as hybrid-mode-family OPO (hOPO). We demonstrate its unprecedented robustness in dispersion versus device geometry, pump frequency, and temperature. Moreover, we show the capability of the hOPO scheme to generate a few milliwatts of output signal power with a power conversion efficiency of approximately 8~$\%$ and without competitive processes. The hOPO scheme is an important counterpoint to existing sOPO approaches, and is particularly promising as a robust method to generate coherent on-chip visible and infrared light sources.
\end{abstract}

\maketitle

\section{Introduction}
On-chip generation of coherent light is critical in the field-level deployment of many applications. Miniaturization of wavelength references~\cite{Hollberg2005}, optical clocks~\cite{Ludlow2015}, and quantum processing elements~\cite{Simon2010} requires on-chip visible lasers that are stabilized to atomic systems and/or used as pump lasers. Aside from developing lasers at the wavelengths of interest onto silicon chips directly~\cite{Liang2010}, nonlinear optics is another route for coherent light generation~\cite{Boyd2008}. While many nonlinear optical mixing processes can generate new colors of light, the output frequency of most processes is \kst{directly linked to the frequency or frequencies} of the input laser(s). For example, second-/third-harmonic generation with pump frequency of $\omega_\text{p}$ only generates light at 2$\omega_\text{p}$/3$\omega_\text{p}$; sum-/difference-frequency generation with inputs of $\omega_{1}$ and $\omega_2$ only generates an output at $\omega_{1}\pm\omega_{2}$. Compared to other processes, optical parametric oscillation (OPO) is unique in enabling coherent light generation at new frequencies without this strict tie to the input laser frequency --- OPO is able to generate signal and idler light with different frequencies, as long as $\omega_\text{s}+\omega_\text{i} =\sum\omega_\text{p}$, where the specific frequencies of $\omega_\text{s}$ and $\omega_\text{i}$ depend on the dispersion engineering to achieve phase matching at those frequencies.

One promising approach is to use degenerately-pumped OPO based on the third-order nonlinearity ($\chi^{(3)}$) in whispering-gallery-mode resonators on chip~\cite{Vahala2004} to extend the reach of existing mature lasers (for example, at 780~nm or 1550~nm) to wavelengths that are otherwise difficult to access~\cite{Lin2008, Sayson2017, Fujii2017,Sayson2019, Fujii2019,Lu2019C, Lu_Optica_2020, Tang2020, Lipson2021}. This concept~\cite{Lin2008} is a natural extension of pioneering work on whispering-gallery-mode $\chi^{(3)}$ OPO~\cite{Vahala2004, Savchenkov2004}, where signal and idler are close to the pump mode in frequency, so that the dispersion requirements are comparatively simpler. $\chi^{(3)}$ OPO shows many desired characteristics for coherent light generation besides device miniaturization, \kst{including producing signal and idler fields} separated by 190~THz at milliwatt-level power thresholds~\cite{Lu2019C} and spectral coverage greater than an octave~\cite{Sayson2019, Fujii2019, Lu_Optica_2020, Lipson2021}.

However, this $\chi^{(3)}$ OPO approach faces challenges in practice. First, this approach typically requires a near-to-zero but normal dispersion for the pump, which requires a fine balance of higher-order dispersion, and is therefore quite sensitive to geometry and pump frequency. While such dispersion sensitivity is the key to its ability to realize very widely tuned output frequencies for a limited amount of pump frequency tuning (for example, a signal tuning $>$~40$\times$ that of the pump tuning was shown in Ref.~\onlinecite{Lu_Optica_2020}), the aforementioned sensitivity translates to requiring the device geometry to be accurate to within a few nanometers when the pump frequency is fixed~\cite{Lu2019C}. Moreover, the dispersion landscape in the pump frequency band (i.e., crossover from anomalous to normal) means that the OPO behavior can dramatically change from one pump mode to the adjacent one (separated by its free spectral range), and in some cases, result in competitive (undesirable) processes~\cite{Lu2019C, Lipson2021}. Second, while close-band $\chi^{(3)}$ OPO can be quite efficient (e.g., 17~\% conversion efficiency from pump to signal~\cite{Vahala2004}), widely-separated $\chi^{(3)}$ OPO typically exhibits much lower conversion efficiencies below 1~\%, \kst{where the signal (idler) conversion efficiency is an on-chip quantity defined as the signal (idler) power in the output waveguide divided by the pump power in the input waveguide.} Such low efficiency is due to two major reasons. The first is related to coupling~\kst{to the access waveguide, as coupling of signal and idler fields that are widely separated in frequency from the pump, while maintaining adequate pump coupling to inject power in the cavity, can be a challenge~\cite{moilleBroadbandResonatorwaveguideCoupling2019a}.} The second reason is related to dispersion, as the close-to-zero dispersion \kst{that enables widely-separated OPO can also support competitive close-to-pump modulation instability processes~\cite{Jordan_arXiv_2021}}.

Therefore, \kst{many of the} challenges of these $\chi^{(3)}$ OPO devices~\cite{Lin2008, Sayson2017, Fujii2017,Sayson2019, Fujii2019,Lu2019C, Lu_Optica_2020, Tang2020, Lipson2021} are linked to the dispersion of the scheme used to realize phase and frequency matching of the pump, signal, and idler modes. \kst{The aforementioned works typically} use whispering gallery modes from the same transverse mode family (with differing azimuthal orders), which we hereafter refer to as the single-mode-family OPO (sOPO). This single-mode-family approach is also used in other $\chi^{(3)}$ processes in microresonators, such as four-wave mixing Bragg scattering~\cite{Li2016} and Kerr comb generation~\cite{Gaeta2019}. On the other hand, nonlinear optical processes such as $\chi^{(2)}$ OPO~\cite{Bruch2019} and second-harmonic and third-harmonic generation~\cite{Levy2011, Surya2018, Lu2021} often use different transverse mode families to realize phase and frequency matching, as the single-mode-family approach is typically not workable unless used in conjunction with quasi-phase matching~\cite{Wang_Optica_2018, Juanjuan_Optica_2019,Nitiss2021}. The basic reason is that $\chi^{(3)}$ OPO involves annihilated pump photons whose frequency is in-between that of generated signal (higher frequency) and idler (lower frequency) photons, and the corresponding phase-matching relationship can be satisfied by modes of the same family, whereas in harmonic generation, for example, it is typically necessary to use a higher-order mode at the higher frequency in order to phase-match with a fundamental mode at the fundamental frequency. In other words, working with a single-mode family is an option but not a necessity for $\chi^{(3)}$ OPO. Indeed, $\chi^{(3)}$ OPO using higher order modes for signal has been observed in millimeter-size crystalline OPO~\cite{Liang_Optica_2015}; however, due to its high \kst{spectral density of modes, OPO does not occur in only a single signal and idler pair (as often desired), but is instead accompanied by various additional frequency components} through other nonlinear mixing processes.

In this work, we demonstrate the use of different transverse mode families for the pump, signal, and idler fields in a \kst{chip-integrated, $\chi^{(3)}$ microresonator} OPO, which we term hybrid-mode- OPO (hereafter hOPO). We discuss the design principles and various configurations for hOPO, and showcase one example in which the involved mode families \kst{exhibit a modal} anti-crossing. The demonstrated hOPO devices have a threshold power of approximately 10~mW, and show unprecedented robustness against geometric variation (up to 500~nm change in ring width), pump frequency tuning ($\approx$ 1:1 ratio of the output signal and idler tuning to the input pump tuning), and temperature tuning (across a temperature range of 40~$^{\circ}$C). By operating \kst{with} the pump band in a regime of large normal dispersion, hOPO is particularly promising for realizing high conversion efficiency from pump to signal, as most competing four-wave mixing mediated processes are suppressed. To that end, we demonstrate a hOPO process with a on-chip power conversion efficiency of $\approx$~8~\%, and with signal output power as high as $\approx$ 5~mW.


\begin{figure*}[t!]
\centering\includegraphics[width=0.85\linewidth]{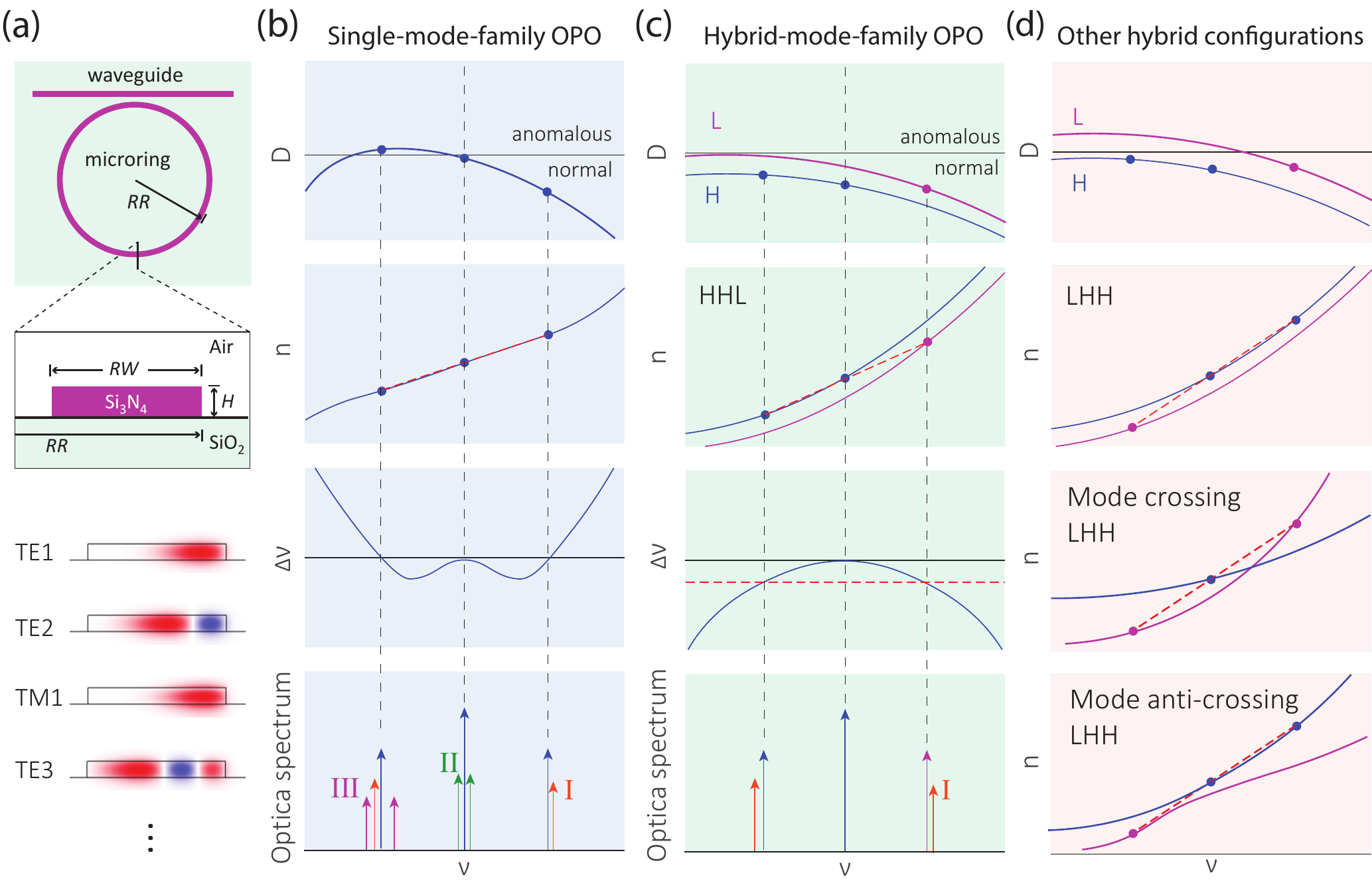}
\caption{\textbf{The physics of hybrid-mode-family optical parametric oscillation (hOPO).} \textbf{(a)} Illustration of a device with a silicon nitride (Si$_3$N$_4$) core with silicon dioxide (SiO$_2$) substrate and air cladding. Its modes are typically either transverse-electric-like (TE) or transverse-magnetic-like (TM). For example, the dominant electric field components of the TE1, TE2, TM1, and TE3 modes at $\approx$ 390~THz (769 nm) in a device with ring radius $RR$~=~28~$\mu$m, ring width $RW$~=~2.8~$\mu$m, and ring thickness $H$~=~323~nm are shown in the bottom of (a), with their effective modal indices sorted from high to low. \textbf{(b)} For single-mode-family OPO (sOPO), the pump operates at a near-to-zero but still normal dispersion (the top panel), and the effective modal indices have to be matched for idler, pump, and signal according to the equation $n_\text{s}/\nu_\text{s}+n_\text{i}/\nu_\text{i}=2n_\text{p}/\nu_\text{p}$ (dashed red line in the second panel) for modes with zero frequency mismatch (third panel) $\Delta\nu = \nu_\text{s}+\nu_\text{i}-2\nu_\text{p}=0$. Such sOPO can have three competitive processes, the process in the designed configuration but with adjacent signal and idler modes also exhibiting OPO (I, shown in red), close-band OPO (II, shown in green) or frequency comb generation (II), and cluster comb generation where many closely-spaced parametric sidebands surround the targeted signal and/or idler modes (III, shown in purple). Further cascaded processes from I-III are not illustrated here. \textbf{(c)} We propose to use hOPO, where two different mode families are used. The pump mode is chosen from the higher-effective-index mode \kst{family} (H, in blue), which exhibits normal dispersion across the entire spectral range under consideration (blue curve in the top panel). The signal can be from the lower-effective-index mode \kst{family} (L, in purple). The matching of effective modal indices (i.e., phase-matching) in this case depends on the difference of the H and L modal indices at the signal wavelength, rather than the dispersion of a single mode family (H or L) as in sOPO. Equivalently, the frequency matching line is shifted downward, due to the difference of effective indices of these two modes at the signal frequency, as shown in the third panel. This configuration is termed HHL, indicating a higher-index (H) or lower-index (L) mode family from which each of the signal, pump, idler (in this order) modes is taken. Because the pump is in a more strongly normal dispersion region than in sOPO, most competitive processes will be excluded and only process I is expected to potentially be present. \textbf{(d)} The hOPO has many other configurations in which it can operate. First, as shown in the top panel, the lower-index mode family (L) can have anomalous dispersion for idler and pump frequencies, with the higher-index mode being normal for all three frequencies; moreover, the dispersion of H and/or L can be shifted up and down. Second, besides the HHL scheme, LHH (or LHL) can also be used as shown in the second panel. Moreover, when two modes are adjacent to each other and similar in effective modal indices, these two modes can exhibit either a direct crossing (third panel) or an anti-crossing (bottom panel).}
\label{Fig1}
\end{figure*}

\section{Nonlinear physics of \lowercase{h}OPO}
The hOPO \kst{scheme can be implemented} in various photonic platforms, similar to \kst{the sOPO scheme}. Here, we use silicon nitride microrings, a popular \kst{device} platform for high-$Q$ $\chi^{(3)}$ nonlinear optics~\cite{Moss2013} including octave-spanning frequency combs~\cite{Okawachi2011, Li2017, Karpov2018}, quantum frequency conversion~\cite{Li2016, Singh2019}, harmonic generation~\cite{Levy2011, Surya2018, Nitiss2021,  Lu2021}, and also sOPO~\cite{Lu2019C, Lu_Optica_2020, Lipson2021}. The device structure we use is illustrated in Fig.~\ref{Fig1}(a), with a stoichiometric silicon nitride (Si$_3$N$_4$) core, silicon dioxide (SiO$_2$) substrate, and air cladding. The microring typically supports transverse-electric-like (TE) modes $\{$TE1, TE2, TE3 ...$\}$ whose dominant electric field component is in the radial direction, and transverse-magnetic-like (TM) modes $\{$TM1, TM2, TM3 ...$\}$, whose dominant electric field component is in the vertical direction. Here, the mode index refers to the number of antinodes in the electric field in the radial direction; though higher-order modes in the vertical direction are also possible depending on the Si$_3$N$_4$ thickness, they do not factor in this work. Modes from these different families are interleaved in frequency space.

The sOPO \kst{scheme} is usually operated with the pump in the normal dispersion regime, to avoid close-band parametric processes (including close-band OPO and Kerr comb generation). Simultaneously, the energy conservation (frequency matching) criterion $\nu_\text{s}+\nu_\text{i}=2\nu_\text{p}$ must be satisfied (to within the cavity linewidth, approximately) for the targeted signal, idler, and pump modes $\nu_\text{s,i,p}$. This necessitates dispersion profiles such as that shown in the top panel of Fig.~\ref{Fig1}(b), where the pump has a close-to-zero dispersion. Here, $D$ is the dispersion parameter given by $D = - \frac{\lambda}{c} \frac{d^2 n}{d\lambda ^2}$, where $c$, $\lambda$, and $n$ represent the speed of light in vacuum, vacuum wavelength, and effective refractive index of the cavity modes, respectively. $D>0$ corresponds to anomalous dispersion and $D<0$ corresponds to normal dispersion. To achieve such a dispersion profile, the device geometry is fixed to an aspect ratio $H:RW$ ($H$ represents the device thickness/height and $RW$ represents the ring width) that depends on the mode and frequency in use, for example, $H:RW~=~1:1.7$ when using TE1 pumped at 780~nm in an air-clad geometry~\cite{Lu_Optica_2020} and $H:RW~=~1:2.7$ when using TM2 pumped at 780~nm in a SiO$_2$-clad geometry~\cite{Lipson2021}. While successful in realizing widely-separated OPO, such designs are generally quite sensitive to geometry, and control of aspect ratio alone is insufficient in dispersion design. Instead, specific parameter sets of $H$ and $RW$ are required, and these parameters are not continuously tunable over a wide range~\cite{Lu2019C}.

Rather than using the $D$ parameter, design using the effective modal index ($n$) across the wavelength range of interest can be helpful [second panel in Fig.~\ref{Fig1}(b)], that is, to engineer $n$ as a function of frequency to satisfy $n_\text{s}/\nu_\text{s}+n_\text{i}/\nu_\text{i}=2n_\text{p}/\nu_\text{p}$ (i.e., phase matching) for $\nu_\text{s}+\nu_\text{i}=2\nu_\text{p}$. This is essentially equivalent to engineering higher-order dispersion around the pump frequency~\cite{Sayson2019, Lipson2021}. The end-result can also be plotted by a frequency mismatch, given by $\Delta\nu = \nu_\text{s}+\nu_\text{i}-2\nu_\text{p}$, assuming $n$ (or equivalently, the azimuthal mode number $m$ in the microring) is matched, with all configurations of $\nu_\text{s}$ and $\nu_\text{i}$ while fixing $\nu_\text{p}$. The typical frequency mismatch curve illustrated in Fig.~\ref{Fig1}(b) shows two zero-crossings, corresponding to the signal and idler frequencies, and a negative mismatch near the pump, indicative of normal dispersion. A more complicated and flatter profile leads to frequency mismatch curves with four zero-crossings, which results in two sets of signal/idler pairs, but is less representative and not shown here~\cite{Lu2019C}.

The basic dispersion profile using the sOPO approach has multiple potential pathways for competitive four-wave mixing mediated processes~\cite{Jordan_arXiv_2021}. This includes other signal/idler pairs that can be frequency-matched with the pump (I), close-band OPO or frequency comb generation, particularly at high power (II), and cluster comb generation for an idler whose surrounding dispersion is anomalous (III). These competitive processes in turn limit the OPO conversion efficiency, particularly when pump power is much higher than threshold~\cite{Lu_Optica_2020, Lipson2021}. While there are likely ways to mitigate these competitive processes, for example, by controlling coupling or using a larger free-spectral range, the potential for such competitive processes is essentially inherent to the dispersion profile in use. Combined with the aforementioned geometric sensitivity of the sOPO approach, it seems clear that while this approach is promising in many respects including the low threshold power and the wide spectral separation between signal and idler, it also possesses built-in challenges that are worth addressing through a different dispersion engineering scheme.

The hOPO approach we present here employs a significantly different dispersion engineering approach to address these challenges. In contrast to the sOPO design with pump frequencies close to zero dispersion, we here use devices with wider ring widths, for example, with an aspect ratio of \kst{$H:RW$~$\approx$~1:8.7} (parameters specified in caption of Fig.~\ref{Fig1}(a)). Such a device has its first four modes (sorted by decreasing effective modal index) as TE1, TE2, TM1, and TE3, as shown in the bottom part of Fig.~\ref{Fig1}(a). Lower-order modes have higher effective modal indices because of better confinement to the Si$_3$N$_4$ core. TM1 appears after TE2 because the microring aspect ratio of 8.7:1 is much larger than the aspect ratios typically used in single-mode-family OPO~\cite{Lu_Optica_2020} or Kerr frequency comb generation~\cite{Li2017} in similar air-clad systems, \kst{as previously noted.} The ring has a large normal dispersion across the entire spectral range under consideration (top panel of Fig.~\ref{Fig1}(c)). The compensation of the effective modal indices (i.e., phase-matching) is not achieved by a fine dispersion design as in the sOPO case, but is rather achieved by choosing a different transverse mode for either the signal (or idler) mode, as compared to the other two remaining modes. For example, in the second panel of Fig.~\ref{Fig1}(c), we show a case in which phase-matching is achieved by using higher-index (H), higher-index (H), and lower-index (L) modes for idler, pump, and signal (frequencies ordered from low to high). In terms of frequency mismatch $\Delta\nu$, the matching of the effective modal indices (dashed red line) is equivalent to matching a dashed line that is shifted in the normal dispersion regime (where $\Delta \nu < 0$), with the shift related to the effective modal index difference of the signal mode in this case. As a result, this hOPO scheme will likely be free of many noise processes (e.g., those labeled as II and III as described earlier), and the main potential competitive process remaining is the adjacent signal and idler modes of the same configuration (i.e., HHL), illustrated by process I in the bottom panel of Fig.~\ref{Fig1}(c)).

\begin{figure*}[t!]
\centering\includegraphics[width=0.76\linewidth]{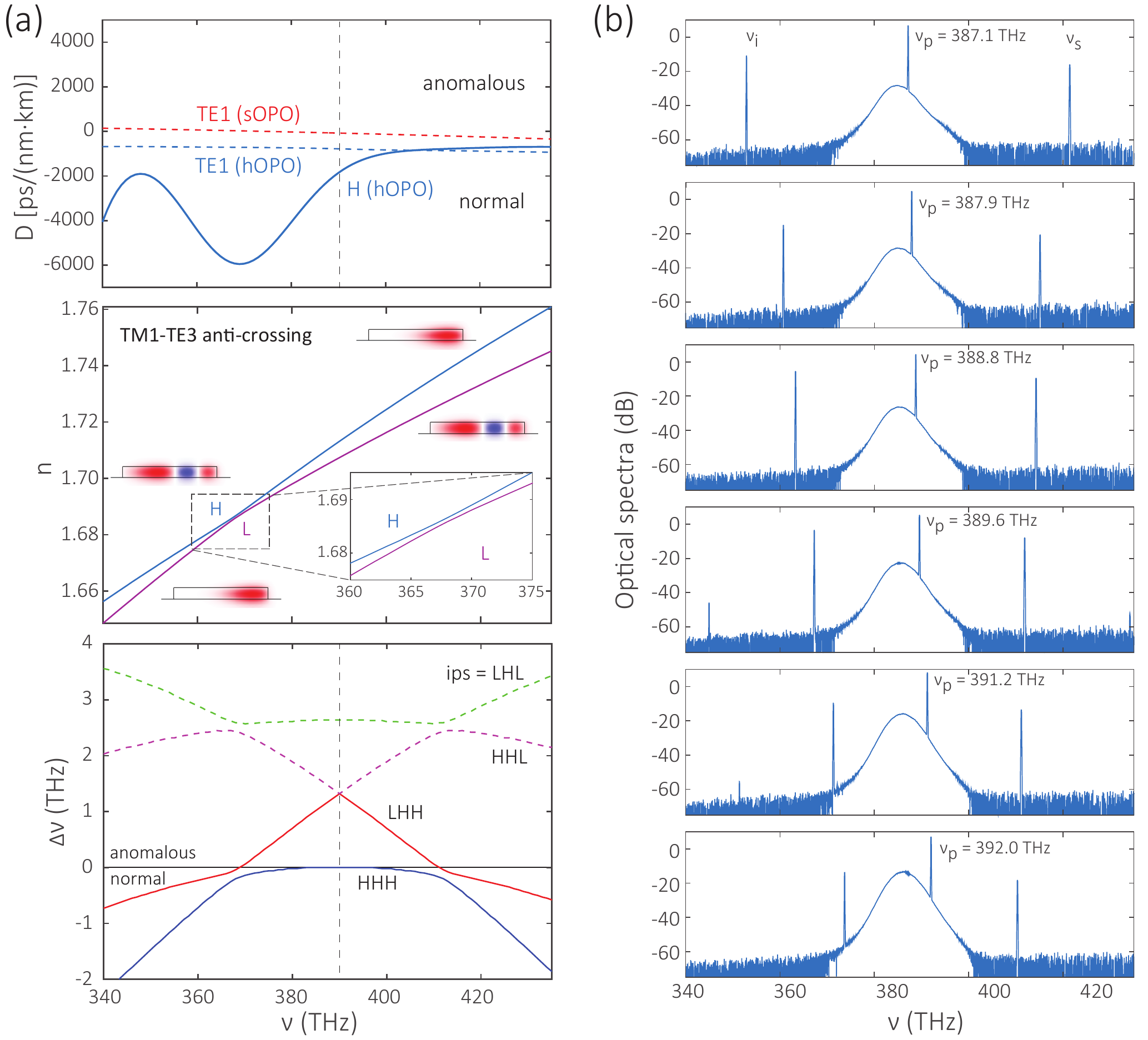}
\caption{\textbf{Demonstration of hOPO.} \textbf{(a)} Simulation of dispersion, effective modal index, and frequency mismatch to support hOPO. The top panel shows the dispersion parameter for the higher \xl{family} (H, blue curve), which hybridizes TM1 and TE3 modes in a microring with $RW$~=~2.8~$\mu$m and $H$~=~323~nm. Its dispersion around 390~THz is -1860 ps/(nm$\cdot$km). In comparison, the dispersion for the TE1 mode (blue dashed line) in such a device is -830 ps/(nm$\cdot$km), and the dispersion around the pump mode in sOPO~\cite{Lu_Optica_2020} (red dashed line, $RW$~=~825~nm) is very close to zero dispersion, i.e., $>$ -100 ps/(nm$\cdot$km). The middle panel shows the effective modal indices of the higher and lower families, which hybridize TM1 and TE3 modes. The curves exhibit an anti-crossing at $\approx$~368~THz, where the higher family (H) shifts from TE3 to TM1, and the lower family (L) shifts from TM1 to TE3, with frequency increased from 340~THz to 435~THz. The bottom panel shows the frequency mismatch ($\Delta\nu$) in all four possible configurations when the pump is at 390~THz and from the H family. The four configurations are labeled based on the chosen families for the idler, pump, and signal modes, and are LHL, HHL, LHH, and HHH. Choosing the pump from the H family ensures that it is in the normal dispersion regime, as shown in the curve for HHH. The only configuration that supports the frequency matching needed for OPO is LHH, with idler at $\approx$~370~THz and signal at $\approx$~410~THz. \textbf{(b)} Experimentally recorded optical spectra in this device as the pump frequency is tuned from 387~THz to 392~THz in approximately one FSR steps. On the y axes, 0 dB is referenced to 1 mW, i.e., dBm.}
\label{Fig2}
\end{figure*}

The hOPO scheme can be \xl{realized} in many other operating configurations. First, as shown in the top panel of Fig.~\ref{Fig1}(d), the lower-index mode family (L) can exhibit anomalous dispersion at the idler and pump \kst{wavelengths}, and the dispersion of H and L can be shifted up and down (subject to keeping the pump, \kst{taken from the H mode family}, in normal dispersion to avoid the close-band OPO/Kerr comb generation on the pump mode family). Second, besides the HHL (high-high-low index) configuration, LHH or LHL can also be used, though the index mismatch of LHL we have observed in our simulations is typically too large to be useful. Finally, when two mode families are adjacent to each other and similar in effective modal indices, they can exhibit a direct crossing or an anti-crossing. From the perspective of hOPO, these two cases are not particularly different as long as the participating modes (i.e., those whose effective modal indices enable phase-matching for frequency-matched modes) are situated away from the crossing/anti-crossing point. In the mode anti-crossing case, an additional benefit is that \kst{reasonable} mode overlap is guaranteed because of mode hybridization. In other hOPO cases (without mode hybridization), adequate field spatial mode overlap for modes from differing families is required, similar to other nonlinear mixing processes using different families, for example, $\chi^{(2)}$ OPO~\cite{Bruch2019} and second-/third-harmonic generation~\cite{Levy2011, Surya2018, Lu2021}.

In the following three sections we demonstrate a hOPO configuration involving TM1 and TE3 mode families that are used in a region of frequency space in which they exhibit an anti-crossing in the idler band. We highlight that this configuration is particularly robust in terms of its tuning with respect to ring width, pump frequency, and temperature. We are able to realize output powers in each of the signal and idler bands of a few milliwatts, with on-chip pump-to-signal (and pump-to-idler) conversion efficiency of about 8~$\%$. 

\begin{figure*}[t!]
\centering\includegraphics[width=0.85\linewidth]{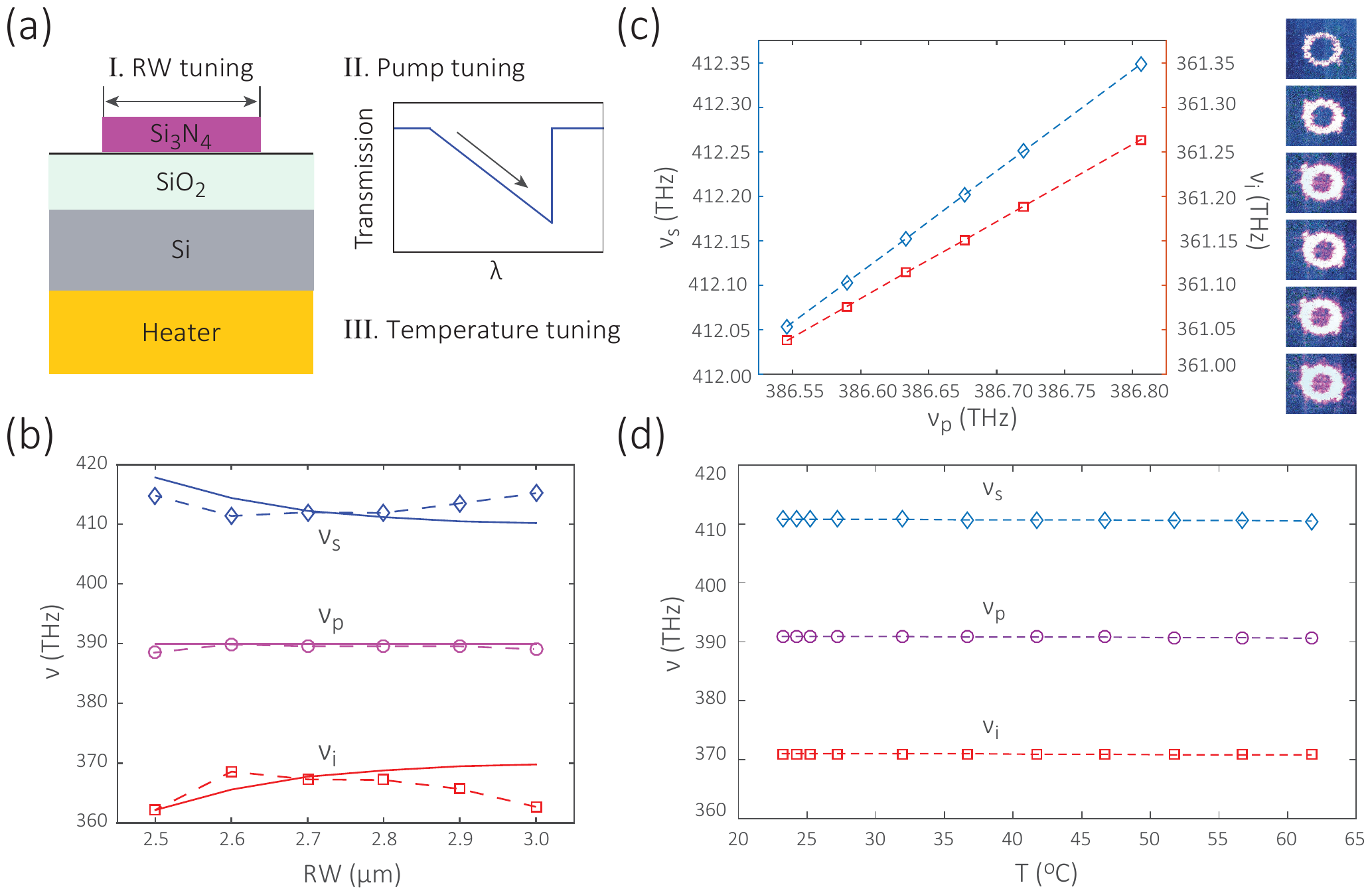}
\caption{\textbf{Robustness of the hOPO output to device geometry, pump detuning, and temperature.}
\textbf{(a)} Illustration of the tuning of ring width ($RW$), pump frequency ($\nu_\text{p}$), and temperature ($T$). Here the $RW$ is changed with the outer ring radius ($RR$) fixed. The pump is tuned from shorter wavelength to longer wavelength, to follow the triangular shape of thermal bistability. \textbf{(b)} hOPO is supported when $RW$ is changed from 2.5~$\mu$m to 3.0~$\mu$m. Diamonds, circles, and squares represent signal, pump, and idler frequencies of the hOPO device. Dashed lines are for guidance. Solid lines are theoretical predictions based on finite-element-method simulation. \textbf{(c)} In the device with $RW$ = 2.6~$\mu$m, when the pump laser is coupled into the cavity with its frequency varying from 386.8~THz (right) to 386.5~THz (left), signal and idler light are tuned 0.3~THz and 0.2~THz, respectively, in a continuous fashion. The microscope images on the right show the signal light scattered by the microring surface roughness, where the scattered light is brighter (from top to bottom), as the pump laser is coupled deeper into the cavity. Pump and idler light are filtered out by a short-pass filter. \textbf{(d)} We carry out temperature tuning from 23~$^\text{o}$C to 62~$^\text{o}$C, in a device with $RW$ = 2.8~$\mu$m, and for a pump mode close to 390~THz. The hOPO output is found to be stable across this temperature range of $\approx$~40~$^\text{o}$C. Uncertainties in the measured data are smaller than the data point size.}
\label{Fig3}
\end{figure*}

\section{Realization of \lowercase{h}OPO}
We start with a device with nominal dimensions of $RR$~=~28~$\mu$m, $H$~=~323~nm, and $RW$~=~2.8~$\mu$m, for which the TM1 and TE3 mode families exhibit a mode anti-crossing near 370~THz, as shown in Fig.~\ref{Fig2}(a). This arrangement resembles the LHH configuration illustrated in the bottom panel of Fig.~\ref{Fig1}(d). The higher-index family (H) changes from TE3 to TM1 \kst{when moving} from low frequency to high frequency, while the lower-index family (L) changes from TM1 to TE3, with the dominant mode profiles for TE3 and TM1 shown in the insets. The mode anti-crossing happen at around 368~THz, where the H-family and the L-family are both approximately 50~\% in TM1 and TE3.

Because these two modes are hybridized, describing them as TM1 or TE3 mode families is not accurate, but distinguishing \kst{the effective index curves on either side of the anticrossing as a higher-index (H) family and a lower-index (L) family is clear}, following the definition from Fig.~\ref{Fig1}. From this effective index behavior in Fig.~\ref{Fig2}(a), we calculate the frequency mismatch ($\Delta\nu$) of four configurations with idler, pump, and signal noted as LHL, HHL, LHH, and HHH. Because of the aspect ratio of the microring cross-section, the pump mode (taken from the H family) is in a region with large normal dispersion, as shown in \kst{the HHH curve}. The only configuration that supports frequency matching for OPO is LHH, with idler at $\approx$~370~THz and signal at $\approx$~410~THz created from a pump at 390~THz. The HHL and LHL configurations are too anomalous ($\Delta\nu>$~1~THz is larger than any potential Kerr shift) to support OPO, while HHH exhibits $\Delta\nu$~$<$~0 throughout, so that the requisite frequency matching is never achieved as any Kerr shifts lead to further frequency mismatch.

\begin{figure*}[t!]
\centering\includegraphics[width=0.85\linewidth]{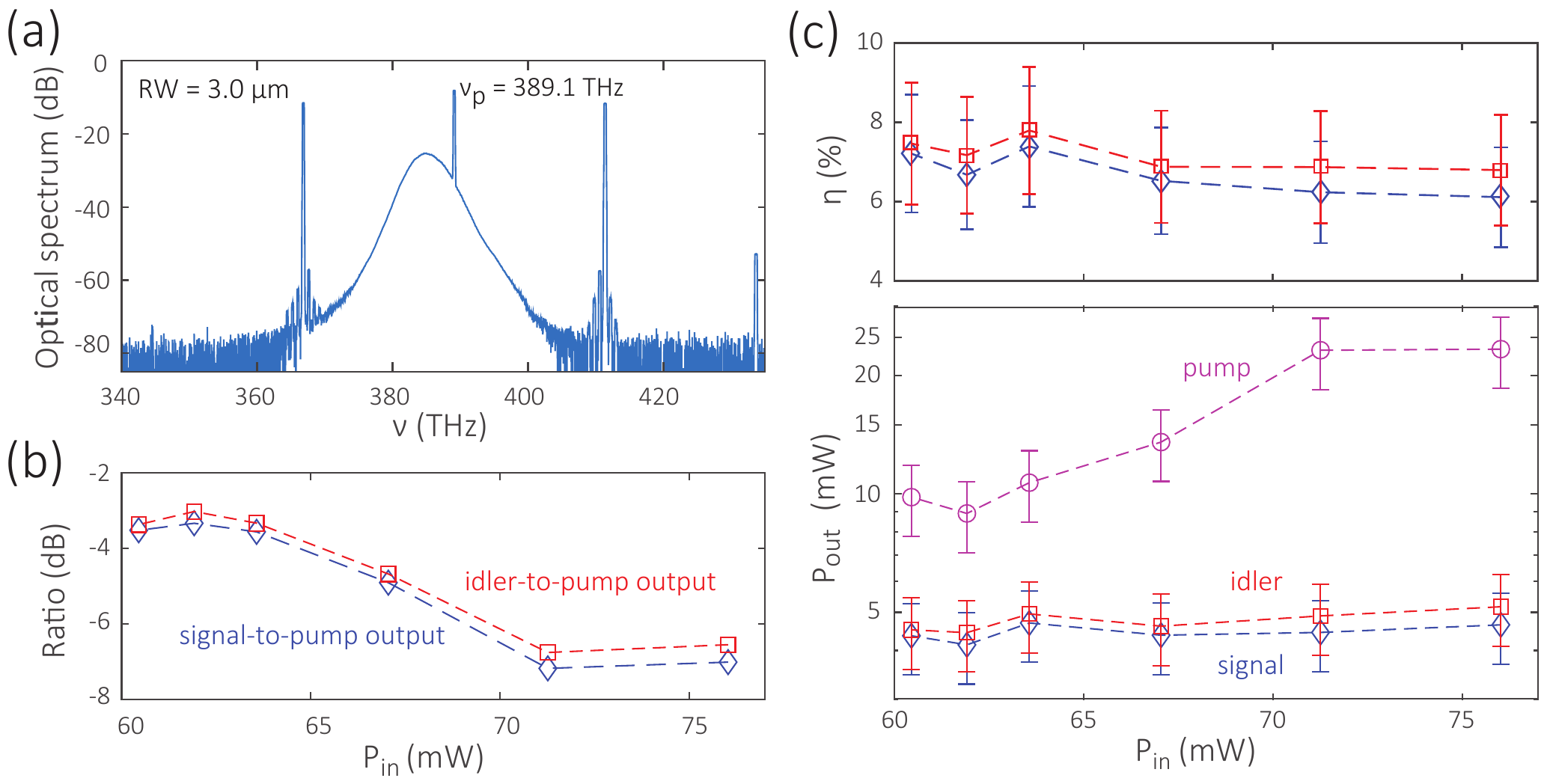}
\caption{\textbf{Efficient conversion and output power from an hOPO device.}
\textbf{(a)} A relatively flat hOPO output optical spectrum from a $RW$~=~3~$\mu$m device pumped at 389.1~THz. Signal and idler outputs are within -3.5~dB of the pump output. On the y axis, 0 dB is referenced to 1 mW, i.e., dBm. \textbf{(b,c)} A power dependence study with the pump frequency fixed. The ratio in (b) is collected from optical spectra as in (a). \textbf{(c)} The efficiency is given by $\eta^{s,i} = P^{s,i}_\text{out}/P_\text{in}$ for signal and idler, respectively, where $P^{s,i}_\text{out}$ is the signal/idler power in the output waveguide and $P_\text{in}$ is the input pump power in the waveguide. The error bars in (c) are $\approx$ 1~dB and estimated by the one-standard deviation uncertainty of the fiber-chip coupling. The uncertainty in (b) is smaller than the data point size.}
\label{Fig4}
\end{figure*}

In Fig.~\ref{Fig2}(b) we record optical spectra measured in this device as the pump frequency is varied from 387~THz to 392~THz in steps of one free spectral range (FSR). hOPO always occurs for various pumping frequencies, in contrast to sOPO where only a few pump modes with the smallest normal dispersion support OPO. The idler and signal frequencies when pumped around 390~THz are in agreement with the simulations to within 3~THz. The threshold power of this hOPO is $\approx$ 10~mW, with Supplementary Information Fig.~S1 showing the evolution of the output spectrum around this power. Given that the resonator quality factors are similar to those observed in previous sOPO work~\cite{Lu2019C}, with the intrinsic quality factors in the range of 1$\times$10$^6$ to 2$\times$10$^6$, the increase of this threshold power (around a factor of 10) is likely caused by a reduction of mode overlap and/or a reduction of $\chi^{(3)}_{1112}$ in comparison to $\chi^{(3)}_{1111}$, but the precise assignment is difficult to isolate in experiment. We also show in Supplementary Information Fig.~S2 another case of hOPO, in this case in a thicker microring ($H$~=~385~nm) for which the mode families involved are TM1 and TE2. We find that the behavior of this hOPO configuration is qualitatively similar to the above case where TM1 and TE3 modes are used.

\section{Robustness of \lowercase{h}OPO}
The physics of hOPO suggests that it should be very robust to device dispersion. As described earlier, the underlying reason is that phase-matching is realized by considering the difference in effective modal index between two mode families at one frequency (signal or idler), instead of through fine balancing of higher-order dispersion in one mode family. A consequence of this is that a change to device dimensions over a wide range should only lead to a small change of signal and idler frequencies, as long as the pump mode is still in the normal dispersion regime. In this section, we carry out studies of how the hOPO device output tunes as a function of ring width ($RW$) variation, pump detuning, and temperature tuning, as illustrated in Fig.~\ref{Fig3}(a).

First, we observe in Fig.~\ref{Fig3}(b) that hOPO is very robust with respect to $RW$, with the output spectrum remaining consistent with the signal and idler frequencies remaining in the same frequency bands when varying $RW$ from 2.5~$\mu$m to 3.0~$\mu$m. We measure the hOPO spectra with $\nu_\text{p}~\approx~$390~THz in Supplementary Information Fig.~S3 and note that signal and idler spectra are clean and free from other \kst{spectral tones}. Such clean and robust hOPO across 500~nm ring width tuning has not been observed previously in sOPO devices; for example, in ref.~\onlinecite{Lu_Optica_2020}, OPO within a much smaller ($\approx$~10~nm) $RW$ range while pumping within the same frequency range as in Fig.~\ref{Fig2}(c) was studied. 
The signal, pump, and idler frequencies are plotted in diamonds, circles, and squares in Fig.~\ref{Fig3}(b), and the simulated results are plotted as solid lines. We can see the experimental results and the theoretical predictions agree well, with a slight deviation of $\approx$ 5~THz on the $RW$~=~3~$\mu$m side that is likely due to imprecise knowledge of the fabricated geometry due to nanofabrication uncertainty.

In comparison to sOPO, where the dispersion is such that the signal and idler frequencies tune widely with small changes to the pump frequency, the hOPO is expected to have a moderate tuning ratio close to 1. We verify this point in experiments shown in Fig.~\ref{Fig3}(c). In the device with $RW$ = 2.6~$\mu$m, when the pump laser is coupled into the cavity with its frequency varied from 386.8~THz (right) to 386.5~THz (left), the generated signal and idler fields are tuned 0.3~THz and 0.2~THz, respectively, in a continuous fashion. This near $1:1$ ratio is quite different from the previous tuning ratio of up to $43:1$ as reported in sOPO~\cite{Lu_Optica_2020}, which suggests that sOPO is advantageous with regards to exhibiting broader spectral tuning, while hOPO might have better spectral stability and robustness to dimensional variation. Moreover, when we carry out temperature tuning over a range of 40~$^\text{o}$C, in a device with $RW$~=~2.8~$\mu$m and $\nu_\text{p}$~=390~THz, the hOPO \kst{output frequencies are} found to be stable (Fig.~\ref{Fig3}(d)). Such robustness to pump detuning and temperature variation makes the hOPO appealing for field deployment outside of laboratory environments.

The robustness of \kst{the hOPO scheme} also indicates that it should be capable of reaching higher output power for signal and idler. To date, sOPO devices typically show -10~dB to -20~dB lower output power at the signal and idler as compared to the pump output power. When signal and idler are very widely separated, the signal output power level is typically further decreased, likely because of coupling (as the high frequency signal tends to be undercoupled in microring-waveguide geoemtries if the pump is critically-coupled~\cite{moilleBroadbandResonatorwaveguideCoupling2019a}). Besides coupling, the low output power is mainly due to competitive processes \kst{that result from the pump being close to zero dispersion}, which limit the OPO conversion efficiency when the power is significantly above threshold~\cite{Jordan_arXiv_2021}. \kst{Given that hOPO can operate with the pump in a region of much stronger normal dispersion, we expect it to limit these competitive processes.} Here we show an example of such performance for a $RW$~=~3~$\mu$m device pumped at 389.1~THz. We find that this hOPO can exhibit a very flat output spectrum in Fig.~\ref{Fig4}(a), where signal and idler outputs are both within -3.5~dB of the pump output. Next, we change the pump power while fixing the pump frequency, and calculate the conversion efficiencies (for signal and idler) and output powers (for pump, idler, signal) in the waveguides before exiting the chip facet. Examining first the output spectra, we find that the idler and signal output powers are up to -3.0~dB and -3.4~dB lower than the remaining pump power, respectively, in Fig.~\ref{Fig4}(b). Next, the on-chip conversion efficiency is up to (8$\pm$2)~\% and (7$\pm$2)~\% for idler and signal, respectively, in the top panel of Fig.~\ref{Fig4}(c). The output power is up to (5$\pm$1)~mW and (4.6$\pm$0.9)~mW for idler and signal, respectively, in the bottom panel of Fig.~\ref{Fig4}(c). We note that the metrics of flatness of the output spectrum, conversion efficiency, and output power are likely optimized at slightly different regions of parameter space, which is important to consider in optimizing the hOPO for practical applications.

\section{Conclusion}
In summary, we propose and demonstrate hOPO as an alternative route to the sOPO that \kst{has been} the prior default scheme for on-chip $\chi^{(3)}$ OPO. We have achieved unprecedented robustness to device geometry, pump power, and temperature using hOPO. Moreover, the dispersion of hOPO is advantageous in achieving higher conversion efficiency, with up to 8~\% pump-to-signal power conversion efficiency on chip. The hOPO is promising for scalable and robust integration of coherent on-chip light sources. An important goal going forward will be to retain the advantageous properties of hOPO demonstrated in this work while broadening its range of accessible wavelengths.\\

\noindent \textbf{Funding.}~Defense Advanced Research Projects Agency (DARPA-LUMOS); National Institute of Standards and Technology (NIST-on-a-chip).\\

\noindent \textbf{Acknowledgement.} F.Z. acknowledges support under the Cooperative Research Agreement between Theiss Research and NIST-PML (70NANB20H174). X.L. and A.R. acknowledge support under the Cooperative Research Agreement between the University of Maryland and NIST-PML (70NANB10H193).\\


\bibliographystyle{osajnl}
\bibliography{hOPO.bib}

\newpage
\onecolumngrid
\renewcommand{\thefigure}{S\arabic{figure}}
\setcounter{figure}{0}
\setcounter{section}{0}
\section*{SUPPLEMENTARY INFORMATION}
\section{Threshold power}
\begin{figure*}[h!]
\centering\includegraphics[width=0.95\linewidth]{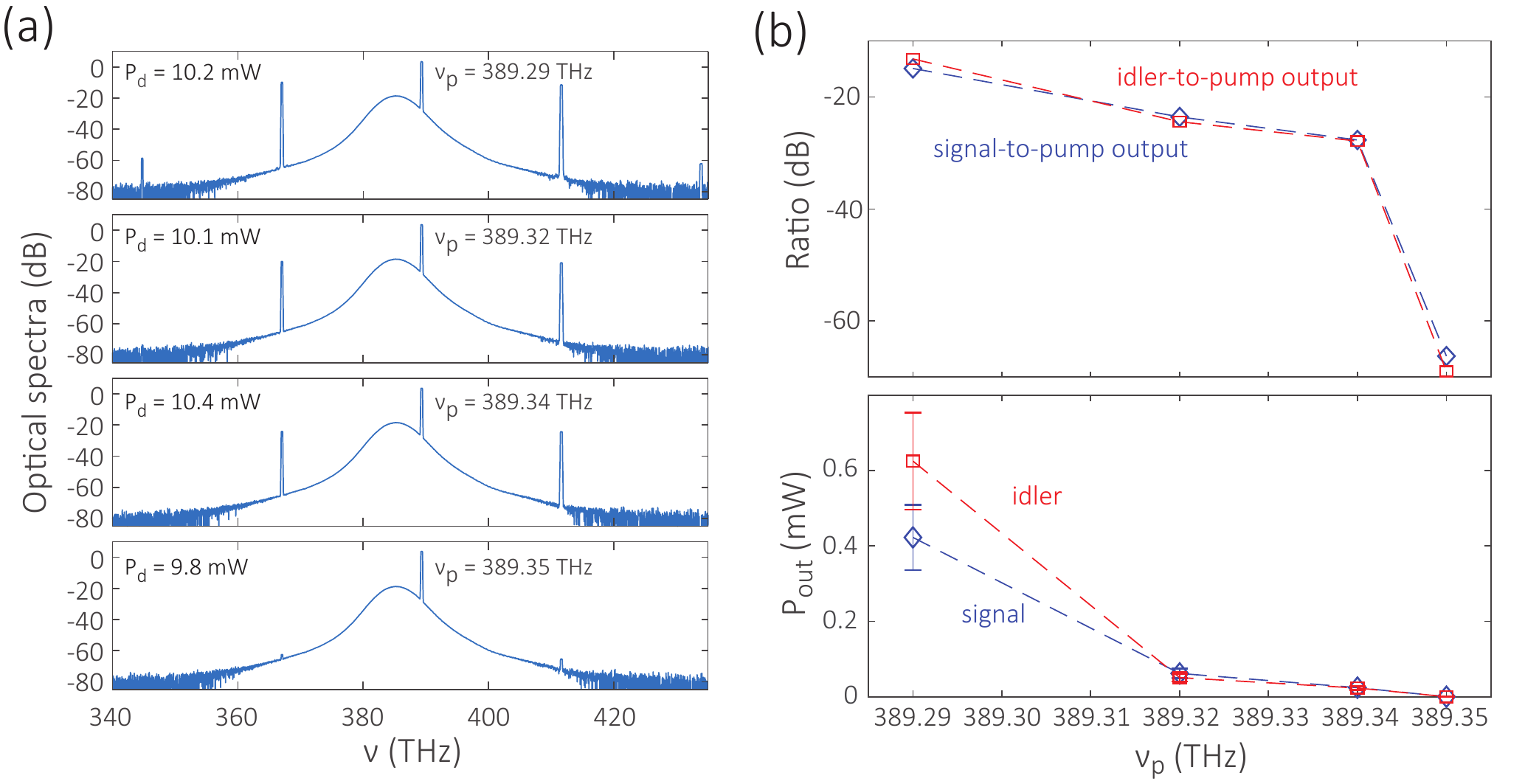}
\caption{\textbf{Threshold power in the hOPO scheme.} \textbf{(a)} In a device with $H$~=~323~nm, $RW$~=~2.8~$\mu$m, and $\nu_\text{p} \approx $389~THz, we carry out a threshold power measurement. At 9.8~mW, OPO is excited but very weak. At 10.24~mW, OPO is fully excited. In the y axes, 0 dB is referenced to 1 mW, i.e., dBm. \textbf{(b)} The relative amount of signal and idler compared to the pump output (top), and the corresponding signal and idler output power on-chip. The error bars in (c) are estimated by the one-standard deviation uncertainty of the fiber-chip coupling. The uncertainty in (b) is smaller than the data point size.}
\label{ESFig1}
\end{figure*}

To estimate the threshold power for the hOPO device described in the main text, we tune the pump laser frequency ($\nu_\text{p}$) into the cavity resonance (Fig.~\ref{ESFig1}(a), from bottom to top) and calibrate the pump power that is dropped into the resonance, given by $P = P_\text{in} (1-T(\nu_\text{p}))$, where $P_\text{in}$ is the input pump power on chip and $T(\nu_{\text{p}})$ is the cavity transmission at the pump frequency. When the pump is tuned below 389.35~THz, and the dropped power is above 10~mW, optical parametric oscillation is excited. Figure~\ref{ESFig1}(b) shows the relative amplitude of signal and idler compared to the pump output (the top panel), and their waveguide-coupled output power (the bottom panel). This threshold power near 10~mW is larger than the 1~mW threshold power observed for single-mode-family OPO within similar Si$_3$N$_4$ microrings~\cite{Lu2019C}, \kst{and is likely} due to a worse mode overlap from using hybrid mode families and a lower optical quality factor for the higher-order mode used here.

\section{A second \lowercase{h}OPO configuration}
\begin{figure*}[h!]
\centering\includegraphics[width=0.95\linewidth]{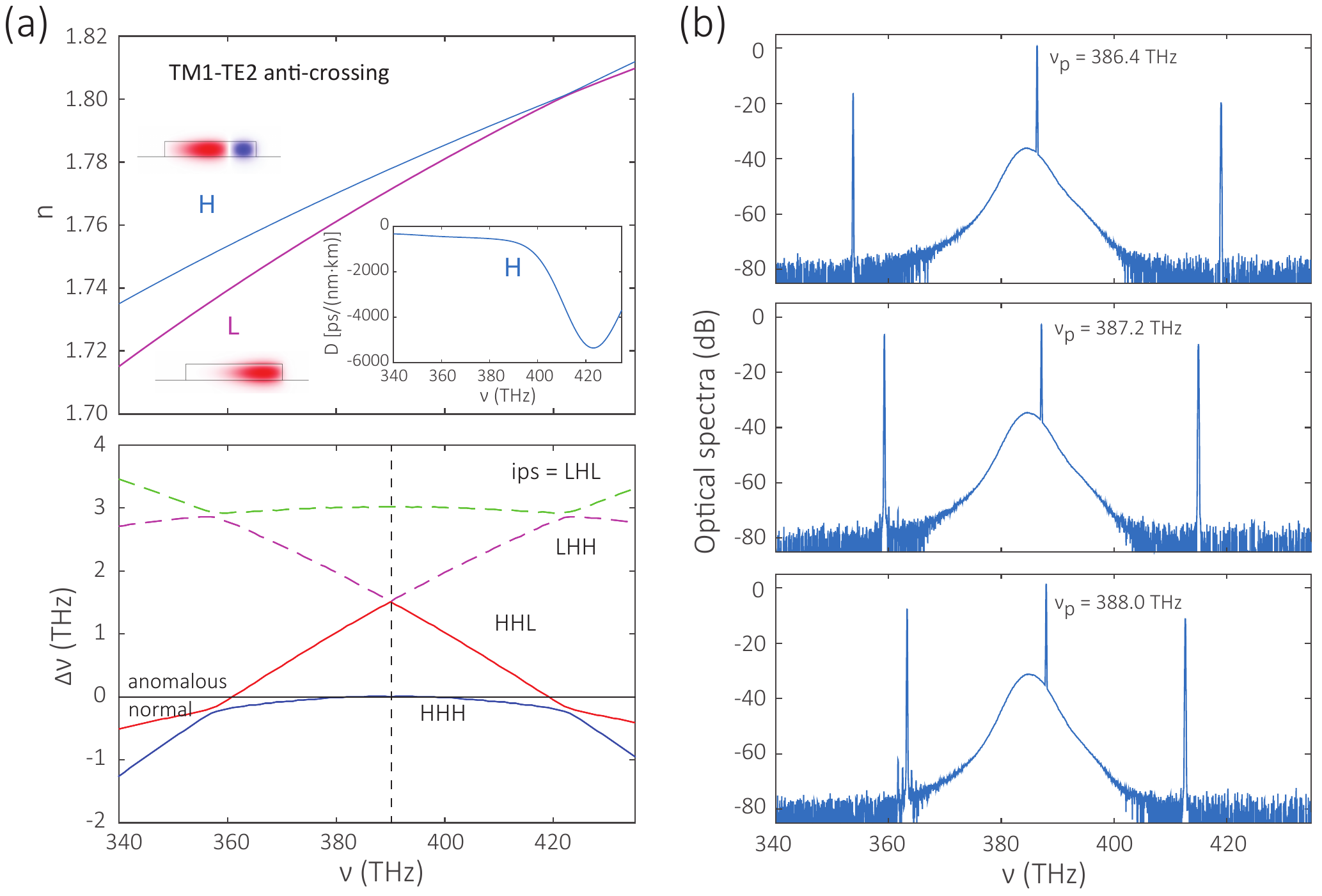}
\caption{\textbf{The hOPO device with TM1-TE2 anti-crossing.} \textbf{(a)} In a device with $H$~=~385~nm and $RW$~=~2.3~$\mu$m, TE2 and TM1 have an anti-crossing around 420~THz, and the devices with pumping on the H family at 390~THz leads to OPO in the HHL configuration. The H mode has a large normal dispersion as shown in the inset. \textbf{(b)} Three experimental hOPO output traces with pump frequency changing from 386~THz to 388~THz.}
\label{ESFig2}
\end{figure*}

We show here another example of hOPO. In this case, the device has a thickness of $H$~=~385~nm and a ring width of $RW$~=~2.3~$\mu$m. As shown in the top panel of Fig.~\ref{ESFig2}(a), the simulation suggests that TE2 and TM1 modes have an anti-crossing around 420~THz, below which TE2 is the higher-index (H) family and TM1 is the lower-index (L) family. The H family exhibits large normal dispersion across the spectral range of interest, as shown in the inset. As shown in the bottom panel, the frequency mismatch ($\Delta \nu$) calculated from the simulation confirms that the H family has a normal dispersion as the HHH curve (blue) is below zero. Out of the four cases with the H mode as the pump, the only case supporting OPO is HHL (red). In experiment, we confirm that such hOPO exists for three consecutive pump modes from 386~THz to 388~THz, as shown in Fig.~\ref{ESFig2}(b). In the 386~THz case (the top panel), the signal is around the mode anti-crossing of 420~THz; in the other two cases, the signals are \kst{at frequencies} below the anti-crossing point. Because the signal is in the L family in the HHL configuration, the signal \kst{in the top panel of Fig.~\ref{ESFig2}(b)} should be about equally split into TM1 and TE2 \kst{(whose mode profiles are shown in the top panel)}, and be more dominantly composed of TM1 rather than TE2 in the two other two panels. In other words, the mode anti-crossing does not seem necessary for hOPO to exist, though it is difficult to confirm or disprove in this case.

\section{Geometric robustness}
\begin{figure*}[h!]
\centering\includegraphics[width=0.95\linewidth]{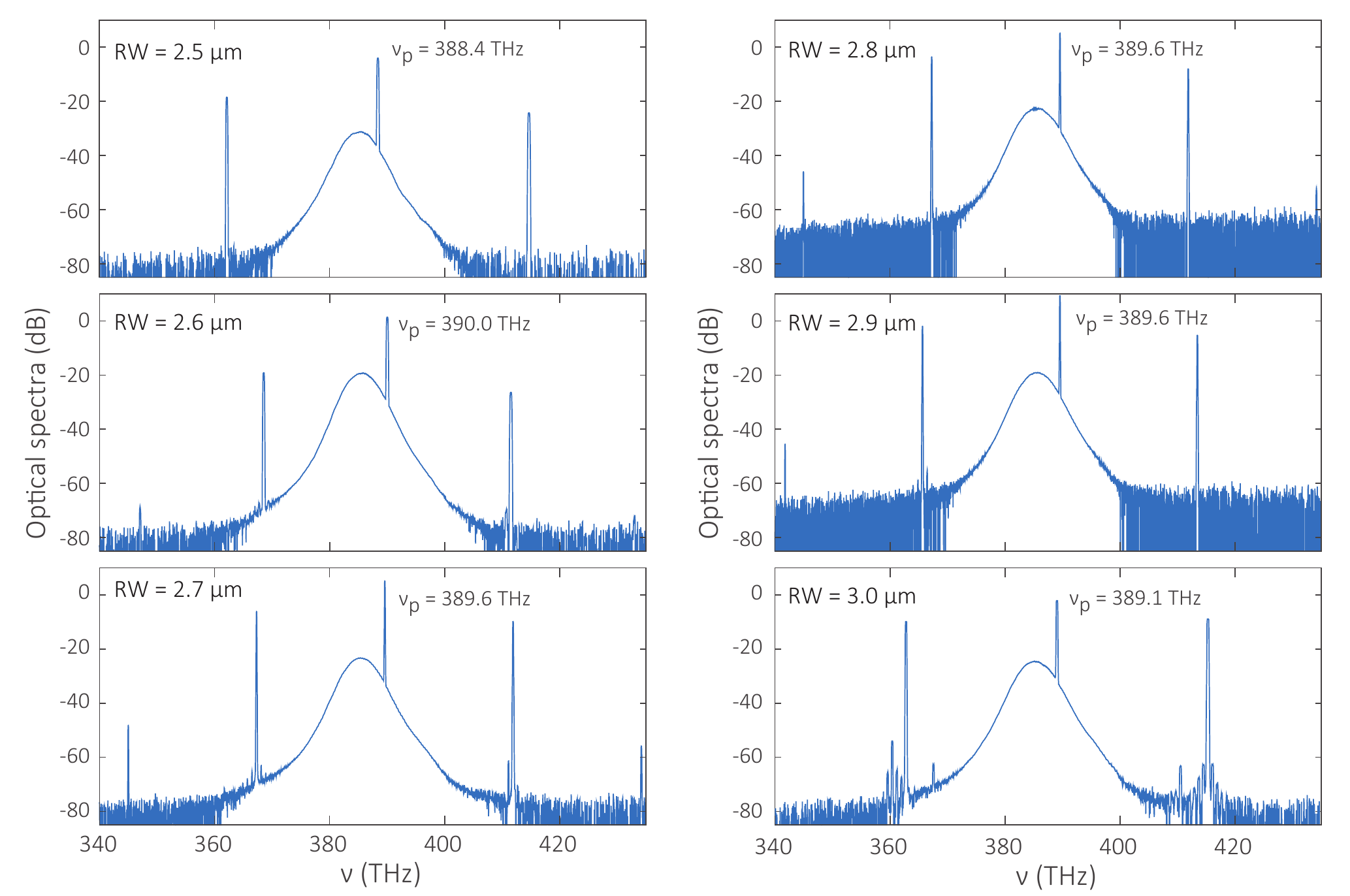}
\caption{\textbf{Geometric robustness of the hOPO output.} When device ring width is varied from 2.5~$\mu$m to 3.0~$\mu$m, hOPO always exists. In the y axes, 0 dB is referenced to 1~mW, i.e., dBm.}
\label{ESFig3}
\end{figure*}

In the main text, we show in Fig.~3(b) that the hOPO exists for different ring widths from 2.5~$\mu$m to 3~$\mu$m with a pump frequency around 390~THz. Here, we show the corresponding optical spectra (Fig.~\ref{ESFig3}) from which the signal, pump, and idler frequencies were recorded.

\end{document}